\documentclass[sigconf]{acmart}
\usepackage[T1]{fontenc}
\usepackage{enumitem}
\usepackage[utf8]{inputenc}
\usepackage{multirow}
\usepackage{textcomp}

\setcopyright{none}

\acmYear{2025} 
\copyrightyear{2025}
\acmConference[FAccT '25]{June 23--26, 2025}{Athens, Greece}

\acmPrice{15.00} % Use the actual price if provided by ACM for final, or a placeholder
\acmISBN{979-8-4007-1482-5/25/06} % Use the actual ISBN or a placeholder like this

\acmDOI{10.1145/3715275.3732059}

\AtBeginDocument{%
  \providecommand\BibTeX{{%
    \normalfont B\kern-0.5em{\scshape i\kern-0.25em b}\kern-0.8em\TeX}}}

\begin{document}

\title{Comparing Apples to Oranges: A Taxonomy for Navigating the Global Landscape of AI Regulation}

\author{Sacha Alanoca}
\email{sachaa@stanford.edu}
\affiliation{%
  \institution{Stanford University}
  \city{Stanford}
  \state{CA}
  \country{USA}
}

\author{Shira Gur-Arieh}
\email{sgurarieh@sjd.law.harvard.edu}
\affiliation{%
  \institution{Harvard University}
  \city{Cambridge}
  \state{MA}
  \country{USA}
}

\author{Tom Zick}
\email{tzick@cyber.harvard.edu}
\affiliation{%
  \institution{Harvard University}
  \city{Cambridge}
  \state{MA}
  \country{USA}
}

\author{Kevin Klyman}
\email{kklyman@stanford.edu}
\affiliation{%
  \institution{Stanford University}
  \city{Stanford}
  \state{CA}
  \country{USA}
}

\renewcommand{\shortauthors}{Alanoca et al.}

\begin{abstract}
AI governance has transitioned from soft law—such as national AI strategies and voluntary guidelines—to binding regulation at an unprecedented pace. This evolution has produced a complex legislative landscape: blurred definitions of “AI regulation” mislead the public and create a false sense of safety; divergent regulatory frameworks risk fragmenting international cooperation; and uneven access to key information heightens the danger of regulatory capture. Clarifying the scope and substance of AI regulation is vital to uphold democratic rights and align international AI efforts. We present a taxonomy to map the global landscape of AI regulation. Our framework targets essential metrics—technology or application-focused rules, horizontal or sectoral regulatory coverage, ex ante or ex post interventions, maturity of the digital legal landscape, enforcement mechanisms, and level of stakeholder participation—to classify the breadth and depth of AI regulation. We apply this framework to five early movers: the European Union’s AI Act, the United States’ Executive Order 14110, Canada’s AI and Data Act, China’s Interim Measures for Generative AI Services, and Brazil’s AI Bill 2338/2023. We further offer an interactive visualization that distills these dense legal texts into accessible insights, highlighting both commonalities and differences. By delineating what qualifies as AI regulation and clarifying each jurisdiction’s approach, our taxonomy reduces legal uncertainty, supports evidence-based policymaking, and lays the groundwork for more inclusive, globally coordinated AI governance.
\end{abstract}




\maketitle

% The permission block and ACM Reference Format you included in your text
% are typically generated automatically by the \maketitle command and the
% \setcopyright, \acmConference, \acmDOI, \acmISBN metadata.
% You should not need to type it manually.

\section{Introduction}
In January 2025, during the first day of the second Trump administration, Biden’s Executive Order for “Safe, Secure, and Trustworthy Development and Use of Artificial Intelligence (AI)” was revoked. Subsequently, a new Executive Order was introduced for “Removing Barriers to American Leadership in AI.” Around the same time, on the international front, South Korea unveiled its Act on the Development of AI and Establishment of Trust (AI Basic Act) while the European Union (E.U.) withdrew its AI Liability Directive. Between new laws and revoked orders, the global landscape of AI regulation is quickly evolving. In this volatile context, there is a pressing need for a taxonomy of AI regulation. A well-constructed taxonomy offers more than just a snapshot; it provides a stable, comparative lens that can endure political changes and help track how regulatory models evolve across jurisdictions. By offering conceptual clarity and a common vocabulary, such a taxonomy supports international coordination and helps navigate a fragmented and fast-changing regulatory environment.

Since 2016, the AI governance landscape has become considerably more defined \cite{ceimia2023framework, levesque2024smoke, halim2023vectors, gsell2024regulating}. AI governance encompasses both soft law (e.g., voluntary guidelines, international agreements, industry standards) and hard law (e.g., regional, federal, state, and municipal laws, executive orders). Initially, soft law was the dominant mode for AI governance. These legally non-binding efforts included national AI strategies, voluntary guidelines for AI system providers, and international initiatives such as the OECD AI Principles, the UNESCO Recommendations on the Ethics of AI, and the G7 Hiroshima AI Process \cite{ceimia2023framework, halim2023vectors, gsell2024regulating, gasser2012interop, gutierrez2024consultation}. In parallel, global AI policy platforms emerged such as the OECD AI Policy Observatory (OECD.AI) and the Global Partnership of AI (GPAI), both launched in 2020. This early progress reached a turning point at the end of 2022 with the widespread adoption of Generative AI \cite{gsell2024regulating}. The rapid development of Generative AI sparked global conversations about AI systems’ transformative potential and societal harms, including misinformation, discrimination, job displacement, and copyright infringement. In response, governments around the world have accelerated efforts to regulate AI. A plethora of AI regulations, both proposed and adopted, emerged and shifted the AI governance focus from soft law to hard law.

The proliferation of AI regulatory announcements makes it hard to navigate this quickly moving landscape. Increasingly, understanding AI regulation amounts to comparing “apples to oranges”, creating novel challenges. First, AI regulation currently entails wide-ranging interpretations. Public and private stakeholders alike leverage the semantic ambiguity of “AI regulation” to describe drastically different efforts. On one end of the spectrum of government-led frameworks is the U.K.’s “pro-innovation approach to AI regulation,” \cite{ukgov2023proinnovation} which does not include legally binding requirements or AI bill proposals. On the other end of the spectrum is the E.U. AI Act \cite{euparliament2024resolution}, which extensively regulates AI systems based on a horizontal, risk-based approach, with financial penalties and sanctions in case of non-compliance \cite{halim2023vectors, gsell2024regulating, dixon2022principled}. While both the U.K. and the E.U. use the same umbrella term, their scope is widely different. Asymmetrical understanding of AI regulation can mislead public expectations by creating a false sense of safety and watering-down regulatory efforts to their lowest common denominator. Second, emerging frameworks adopt different AI regulatory strategies. Countries such as China and the U.S. focus on technology-based coverage (e.g., Generative AI), while the E.U., Canada and Brazil embrace a hybrid model targeting both applications (e.g., healthcare) and technology. Different AI regulatory approaches can challenge legal interoperability and fragment international AI coordination \cite{gasser2017legal, gasser2012interop}. Third, opaque understanding of what AI regulation entails can hinder access to participatory governance. The conceptual ambiguity behind “AI regulation” coupled with the quickly moving AI regulation landscape may deter stakeholders, especially from civil society, from participating in the legislative process. This may consequently increase risks of regulatory capture \cite{levesque2024smoke, gsell2024regulating}.

There is an urgent need to address these challenges by clarifying the scope and substance of AI regulation. Scholarship in this realm has primarily focused on the substance of each regulatory framework (e.g., which systems are prohibited under the E.U. AI Act) rather than the meta-structure (e.g., which common regulatory patterns emerge across jurisdictions) \cite{levesque2024smoke, halim2023vectors, gsell2024regulating, dixon2022principled, hine2022american, radu2021steering, roberts2022governing, roberts2022achieving, gutierrez2024consultation}. While in-depth analysis of each AI regulation provides a detailed snapshot at a given time, the rapidly evolving AI regulatory landscape quickly makes any static work obsolete. The ephemeral nature of the space requires an analytical tool to parse through key differences, while drawing comparisons between jurisdictions. Responsive to these common pitfalls, we developed a taxonomy to identify areas of global alignment and potential blind spots. By doing so, our research moves beyond case-by-case analysis to introduce a framework that enables systematic comparison of AI regulations across time and jurisdictions. Such tools can equip diverse stakeholders, ranging from governments, civil society, trade unions, and companies, to grasp long-standing policy and legal analysis. Moreover, by applying our taxonomy to AI regulations stemming from China and Brazil, our analysis goes beyond the traditional Western-centric lens to provide a global overview of the emerging landscape.

In this paper, we introduce a taxonomy, comparative analysis, and data visualization to illustrate emerging trends in AI regulation. Our framework is informed by interviews with regulators and independent experts across each considered jurisdiction. In Section 2, we discuss the challenges posed by the semantic ambiguity of “AI regulation” and the proliferation of diverse regulatory approaches. We clarify this conceptual ambiguity by distinguishing between soft law and legally binding measures and demonstrating how regulatory capture can arise when voluntary commitments are misrepresented as robust legal protections. In Section 3, we provide an overview of existing literature on AI regulatory policy and comparative studies, delineating the gap that our research intends to meet. In Sections 4 and 5, we detail the methodology and taxonomy criteria selection. We emphasize the collaborative feedback loop with policymakers and legal experts to select our eleven-taxonomy metrics: adoption status, novelty, maturity of the digital legal landscape, reach, enforcement, sanctions, operationalization, international cooperation, stakeholder consultation, regulatory approach, and regulatory focus. We further describe the iterative process through which we distilled the taxonomy into accessible data visualizations. Lastly, in Section 6, we apply the taxonomy to five early-mover jurisdictions—the E.U. AI Act, the U.S. AI Executive Order 14110, Canada’s AI and Data Act, China’s Interim Measures for the Management of Generative AI Services, and Brazil’s AI Bill No. 2338/23—to demonstrate its ability to capture a quickly evolving landscape. These case studies were chosen based on their legally binding requirements, government-led nature, global influence, and representation of diverse regulatory approaches. We complete the taxonomy with data visualizations, which translate complex legal texts into actionable insights, fostering participatory AI governance by democratizing access to otherwise dense information \cite{fjeld2020principled}.

\section{Beyond Regulatory Fragmentation}
\label{sec:beyond_fragmentation}
What is AI regulation? Since 2022, regulatory announcements have surged in response to Generative AI developments. The end of 2023 culminated in several milestones. A series of soft law initiatives were launched such as the U.K. Bletchley Declaration, the UN AI Advisory Board, the E.U. AI Office, the U.K. AI Safety Institute\footnote{The UK AI Safety Institute (AISI) has since been renamed AI Security Institute.}, and the G7 Hiroshima AI Process \cite{ceimia2023framework, halim2023vectors, gsell2024regulating}. These soft law efforts were complemented by hard law initiatives, which distinguished themselves by their government-led, legally binding nature (e.g., non-compliance may result in financial penalties or cessation of activity). Significant hard law efforts released in 2023 include China’s Generative AI Interim Measures, the U.S. AI Executive Order 14110, and the E.U. AI Act compromise text \cite{gsell2024regulating, hine2022american, roberts2022governing}.

However, the loose qualification of “AI regulation” to describe both soft law and hard law efforts continues to blur the lines between legally binding and voluntary requirements. Clarifying the semantic ambiguity of “AI regulation” is critical to prevent misalignment and unmet public expectations. If soft law efforts are misperceived as hard law, citizens may falsely believe they are protected by stricter rules, potentially leading to misplaced trust in AI products. Differentiating between legally binding and voluntary guidelines also ensures that regulatory standards are not diluted to the lowest common denominator.

A central motivation behind our project is to democratize access to AI governance. Aligned with FAccT community principles \cite{facctcommunity2024statement} and prior research \cite{levesque2024smoke, birhane2022power}, our work aims to lower barriers to entry to AI governance by clearly delineating soft law from hard law efforts and providing a digestible overview of the global landscape of AI regulation. By translating complex legal frameworks into accessible outputs through data visualizations, we equip stakeholders—including civil society organizations, trade unions, policymakers, legislators and companies—with a nuanced understanding of AI regulation for meaningful participatory governance. This shared foundation helps ensure diverse voices are represented at the decision-making table, mitigating the risks of regulatory capture \cite{levesque2024smoke, gsell2024regulating}.

\section{Related Work}
\label{sec:related_work}

\subsection{Literature Reviews of AI Governance Frameworks}
\label{subsec:lit_reviews}
Our project builds on and extends a growing body of work that analyzes the regulatory AI landscape. Broadly speaking, prior efforts fall into three overlapping categories: descriptive policy trackers, comparative legal analyses, and broad typological frameworks. 

First, descriptive reviews such as the OECD’s AI Policy Observatory (OECD.AI) tracker \cite{oecd2024observatory} offer a comprehensive inventory of national AI strategies, policies, and laws. Likewise, G’sell’s \textit{Regulating under Uncertainty} \cite{gsell2024regulating} offers one of the most detailed accounts to date, identifying co-regulation and enforcement models across the U.S., E.U., Canada, Brazil, and China. These tools are invaluable for mapping the breadth of global activity, but they tend to prioritize detailed coverage over the meta-conceptual structure. 

Second, comparative studies \cite{levesque2024smoke, dixon2022principled, hine2022american, radu2021steering, roberts2022governing, roberts2022achieving} provide deeper analysis of national frameworks, often focusing on transatlantic differences. Halim and Gasser \cite{halim2023vectors} examine diverging regulatory logics—industry self-assessment in the U.S. versus risk-based oversight in the E.U.—while Lévesque \cite{levesque2024smoke} highlights how legislative gaps and corporate influence shape regulatory outcomes in the U.S., E.U., and Canada. These works inform several of the dimensions we track in our taxonomy, particularly around stakeholder involvement and enforcement mechanisms. 

Third, conceptual frameworks such as CEIMIA’s seven-category model \cite{ceimia2023framework} attempt to structure the space by classifying regulatory tools and institutional approaches. However, these frameworks often blur the distinction between hard law and soft law, treating non-binding ethical guidelines alongside legally enforceable instruments. This conflation risks overstating the level of regulatory protection and underestimating the divergence in legal authority across jurisdictions. 

Our project addresses these limitations, while building on this prior work, in three ways. First, we focus exclusively on binding legal instruments, allowing for a clearer comparison of enforceable governance mechanisms. Second, we expand beyond the typical U.S.-E.U. axis by incorporating regulatory efforts from countries like Brazil and China, which reflect alternative priorities that are often absent from Western-centric accounts. Third, we integrate legal analysis with visual design, translating complex legislative texts into accessible formats. Inspired by works like the \textit{Principled Artificial Intelligence} report \cite{fjeld2020principled} and \textit{Calculating Empires} \cite{calculatingempires2024geopolitics}, our data visualizations make regulatory structures legible to a wider range of stakeholders, including policymakers, auditors, and civil society organizations. 

In sum, while previous work has mapped and categorized AI governance efforts, our project contributes a more legally precise, globally inclusive, and publicly accessible framework for understanding and comparing binding AI regulations.

\section{Methodology}
\label{sec:methodology}
This study provides a comprehensive comparative analysis of AI regulatory frameworks across five key jurisdictions: the United States (Executive Order 14110 on Safe, Secure, and Trustworthy Development and Use of Artificial Intelligence \cite{usexecutive2023safe}), the European Union (E.U. Artificial Intelligence Act \cite{euparliament2024resolution}), Canada (Canada’s AI and Data Act \cite{canadagovAIDACompanion}), China (China’s Interim Measures for the Management of Generative AI Services \cite{pwc2023china}), and Brazil (AI Bill No. 2338/23 \cite{dataprivacybrazil2023legislation}).

\subsection{Overview}
\label{subsec:overview_methodology}
Our process began by identifying and collecting legislative documents from each jurisdiction. We prioritized documents with significant regulatory weight and maturity, focusing on frameworks that established enforceable obligations rather than aspirational policy statements or voluntary guidelines.

After the initial selection, we conducted consultations and interviews with experts in AI policy and law. This phase involved engaging with practitioners directly involved in the drafting and implementation of AI regulation, as well as legal scholars and policymakers with expertise in AI governance. These engagements provided valuable insights into the nuances of each framework, shedding light on the policy, economic, and geo-political contexts that shaped their development. Expert input not only validated the accuracy of legislative interpretations but also revealed emerging regulatory trends and areas requiring deeper investigation.

A central component of the research was the development of a taxonomy to systematically compare legislative frameworks across jurisdictions. The taxonomy was iteratively refined through collaborative discussions and expert feedback to ensure it captured critical dimensions of AI regulation. Key areas of focus are legal maturity of the frameworks, the mechanisms established for enforcement, the nature of the regulatory approach, and the degree of stakeholder engagement. The taxonomy served as the foundation for encoding the legislative documents, allowing the research team to conduct structured and consistent comparisons.

Throughout the analysis phase, legislative documents were carefully reviewed and encoded based on the taxonomy. Regular internal review meetings ensured alignment across the research team, allowing for the resolution of discrepancies and enhancing the consistency of the encoded data. The iterative nature of this process allowed for ongoing refinement, ensuring that emerging legislative developments and amendments were incorporated in real-time.

In the final stage, findings were synthesized and translated into visual representations designed to convey complex regulatory insights in an accessible format. The research team collaborated with designers to create data visualizations that effectively highlighted key patterns and divergences in AI governance frameworks. These visualizations were intended to serve as practical tools for policymakers, industry leaders, researchers, and civil society enabling broader engagement with comparative analysis and facilitating more informed decision-making.

\subsection{Selection Criteria for AI Regulatory Frameworks}
\label{subsec:selection_criteria}
The selection of legislative frameworks for inclusion in this study was grounded in a set of well-defined criteria designed to ensure a comprehensive and representative analysis of AI regulatory approaches across jurisdictions. The criteria aimed to prioritize regulatory maturity, diversity of approaches, and global influence, thereby capturing a wide spectrum of AI regulatory strategies.

A primary criterion was the regulatory significance and maturity of the legislative framework. Jurisdictions with established and mature AI regulatory ecosystems were prioritized to provide insights into frameworks that are either currently in force or nearing implementation. This approach aimed to distinguish “regulation”—a specific, enforceable rule created by a government agency to implement a law—from broader policy, which typically outlines intent or guiding principles and serves as a foundation for developing regulations. In essence, regulation represents the concrete, actionable implementation of policy, specifying how to achieve particular goals within a policy area \cite{baldwin2011understanding}. Although the line between policy and regulation is often blurred—policy documents can sometimes include enforceable elements—we prioritized materials that aligned more closely with formal regulations. From such a standpoint, we aimed to clearly delineate between soft law and hard law efforts, where AI regulations only pertained to the hard law realm.

Another essential factor guiding selection was the diversity of regulatory approaches. The study deliberately included jurisdictions that exemplify a broad range of regulatory strategies, from those that establish precise legal mandates (e.g., comprehensive AI bills) to those characterized by more flexible requirements (e.g., executive orders). However, the selection excluded jurisdictions where AI governance is limited to high-level principles or non-binding guidelines. This diversity allowed for a richer comparative analysis, providing insights into the varied methods countries employ to regulate AI technologies.

Global influence constituted a third core criterion. The study emphasized the inclusion of jurisdictions whose AI regulatory frameworks hold substantial sway over international AI governance discourse and possess the potential to shape emerging global standards. By incorporating regions with far-reaching regulatory influence, the study ensured that its findings would resonate broadly and contribute to the ongoing development of international AI governance norms. Based on these criteria, the following legislative documents were selected: the United States AI Executive Order\footnote{We included Biden’s 14110 Executive Order because, despite not being a traditional regulatory document, it introduces enforceable provisions with significant regulatory weight—most notably the FLOPs reporting requirement. This mandate compels developers of advanced AI systems to report computational resources (measured in FLOPs) used in training large AI models, effectively creating binding obligations for compliance.}, the E.U. AI Act \cite{euparliament2024resolution}, Canada’s AI and Data Act \cite{canadagovAIDACompanion}, China’s Interim Measures for the Management of Generative AI Services \cite{pwc2023china}, and Brazil’s AI Bill No. 2338/23 \cite{dataprivacybrazil2023legislation}. These frameworks represent a diverse array of regulatory approaches and leading edge of AI governance globally.

\section{Taxonomy Development and Selection Criteria}
\label{sec:taxonomy_development}
The development of the taxonomy was guided by the following key considerations to ensure comprehensive, objective, and practical analysis across jurisdictions:
\begin{enumerate}
    \item Key Legislative Elements: We identified essential elements within each jurisdiction’s AI legislation to highlight core aspects of their regulatory approaches.
    \item Commonalities Across Documents: The taxonomy incorporates elements consistently present in all legislative documents.
    \item Measurability: Only elements that can be easily measured and assessed without excessive discretion were included. This criterion ensures reliability, consistency, and minimizes the potential for bias during analysis.
    \item Expert Validation: The taxonomy was iteratively refined through expert consultations and repeated cycles of analysis. Expert feedback ensured the robustness and accuracy of the criteria, effectively capturing the nuances and complexities of the legislation.
\end{enumerate}

These selected metrics are not intended to exhaustively capture every aspect of the legislative documents. Instead, they reflect criteria that align with the outlined principles and enable meaningful comparative analysis.

\subsection{Selected Metrics}
\label{subsec:selected_metrics}
\begin{enumerate}[label=\alph*.]
    \item \textbf{Status:} Tracks whether AI legislation has been formally adopted as law yet.
    \item \textbf{Novelty:} This metric assesses whether the legislation is specifically developed to govern AI (new legislation) or if it adapts existing legal frameworks to apply to AI systems. The goal is to distinguish between laws crafted with AI in mind and those repurposed to address AI-related concerns.
    \item \textbf{Maturity of the Digital Legal Landscape:} Maturity assesses the overall advancement and robustness of a jurisdiction's digital and AI governance framework. Jurisdictions with mature digital regulatory landscapes typically demonstrate comprehensive AI strategies and AI-adjacent regulations (e.g., data protection, platform content moderation).
    \item \textbf{Reach:} Identifies sectors subject to AI regulations:
        \begin{enumerate}[label=\arabic*)] % Corrected
            \item Industry: Regulations targeting private sector companies and their AI applications.
            \item Government/Federal Agencies: Regulations guiding AI use within public sector entities.
            \item Individuals/Citizens: Regulations addressing the use of AI systems by individuals.
        \end{enumerate}
    \item \textbf{Enforcement:} This metric focuses on the mechanisms and processes used to monitor, verify, and enforce compliance with AI regulations.
        \begin{enumerate}[label=\arabic*)] % Corrected
            \item Continued Reporting: Mandates periodic updates or assessments to monitor compliance over time.
            \item Third-Party Audits: Independent assessments conducted by external organizations to evaluate and verify compliance with certain standards.
            \item Existing Agency or New Agency: Reflects whether enforcement is handled by existing regulatory bodies or if new agencies are established specifically for AI oversight.
            \item Liability: This metric examines whether actors bear legal responsibility when AI systems cause harm or fail to comply with regulations.
            \item Regulatory Sandboxes: Controlled environments allowing companies to test AI systems under regulatory supervision.
        \end{enumerate}
    \item \textbf{Sanctions:} This category outlines the legal powers granted to enforcement bodies to ensure compliance with AI regulations and impose penalties on violators.
        \begin{enumerate}[label=\arabic*)] % Corrected
            \item Criminal Charges: Frameworks that enable the prosecution of individuals, organizations, or entities for unlawful actions, negligence, or harm caused by AI systems.
            \item Fines: Financial penalties imposed on non-compliant actors for violating AI regulations.
            \item Temporary Injunctions: Legal orders that temporarily halt the development, deployment, or continued operation of AI systems suspected of posing significant risks or violating regulations.
        \end{enumerate}
    \item \textbf{Operationalization:}
        \begin{enumerate}[label=\arabic*)] % Corrected
            \item Standard Setting: Establishes technical and procedural benchmarks for AI development and deployment.
            \item Auditing: Internal or external audits evaluate AI system performance and adherence to standards.
            \item Technical Expertise: Highlights whether frameworks mandate and have the means to include technical expertise within regulatory bodies, ensuring informed oversight.
        \end{enumerate}
    \item \textbf{International Cooperation:} This metric evaluates alignment with global frameworks (e.g., OECD AI principles), which reflects efforts and the willingness of jurisdictions to contribute to international AI governance norms.
    \item \textbf{Stakeholder Consultation:}
        \begin{enumerate}[label=\arabic*)] % Corrected
            \item Civil Society Inclusion: Reflects engagement with non-governmental organizations, ensuring public interest is represented.
            \item Private Sector Inclusion: Captures whether industry stakeholders contribute to the legislative process.
        \end{enumerate}
    \item \textbf{Regulatory Approach:}
        \begin{enumerate}[label=\arabic*)] % Corrected
            \item Ex Ante: A preventive approach that addresses risks before they arise, emphasizing measures like pre-market approval to ensure AI systems meet safety and compliance standards in advance.
            \item Ex Post: A reactive approach that evaluates AI systems after deployment, focusing on retrospective assessments and legal remedies. It allows individuals affected by AI to pursue recourse, including the right to file lawsuits.
        \end{enumerate}
    \item \textbf{Regulation Layer:}
        \begin{enumerate}[label=\arabic*)] % Corrected
            \item Technology-Focused: Regulations target AI infrastructure (e.g., dual-use foundation models).
            \item Application-Focused: Legislative focus on specific AI use cases (e.g., social scoring). This provides tailored oversight for distinct risk categories and applications.
        \end{enumerate}
\end{enumerate}

\subsection{Challenges and Limitations}
\label{subsec:challenges_limitations}
Throughout this study, we encountered several challenges that required thoughtful consideration and strategic solutions to ensure the validity and reliability of our findings. The key challenges and corresponding solutions are outlined below:
\begin{enumerate}
    \item \textbf{Data Availability and Reliability:} One of the primary challenges was the availability and reliability of information related to our taxonomy, particularly metrics that required data not found within the legislation itself, such as budget allocations for regulatory implementation and enforcement, and degree of participation of public and private stakeholders within the legislative process. Moreover, depending on whether sources were issued by people directly involved within the AI legislation or by independent experts, information could represent widely varying or conflicting views. To address these challenges, we cross-referenced multiple sources and consulted with a wide range of subject matter experts, both directly and indirectly involved in the legislative process, to confirm the accuracy of the information and of our evaluation.
    \item \textbf{Fast-Changing Pace of AI Legislation Landscape:} The dynamic nature of AI regulation, characterized by frequent updates, rescinded orders, and new bill proposals, posed a challenge in keeping our analysis current. To mitigate this, we adopted a continuous monitoring strategy to regularly interview legislators and review official government websites, regulatory bodies, and trusted databases to ensure access to the most up-to-date legislative texts.
    \item \textbf{Taxonomy Prioritization for Data Visualizations:} To optimize our taxonomy to illuminate the developments in emerging AI regulations, we undertook an iterative approach to scoping. We drew a temporal line in 2000 and analyzed technology-focused regulations to draw out key metrics. We then interviewed experts to validate and refine our research scope. Finally, we pursued an iterative design process, collaborating with graphic designers and incorporating multiple rounds of feedback to refine visualizations that effectively translated our key findings.
    \item \textbf{Ensuring Uniformity in Criteria Ratings:} Developing a consistent and reliable rating system for the taxonomy criteria was a complex task. To ensure uniform application of these ratings across different documents and team members, we focused on aligning interpretations and refining criteria during our research process, enhancing the robustness and accuracy of our comparative analysis.
\end{enumerate}

\begin{figure*}[htbp]
  \centering
  % Try a fraction of \textheight. 0.7\textheight or 0.75\textheight is often a good start.
  % Adjust this value until the figure fits well with its caption.
  \includegraphics[height=0.95\textheight, width=\textwidth, keepaspectratio]{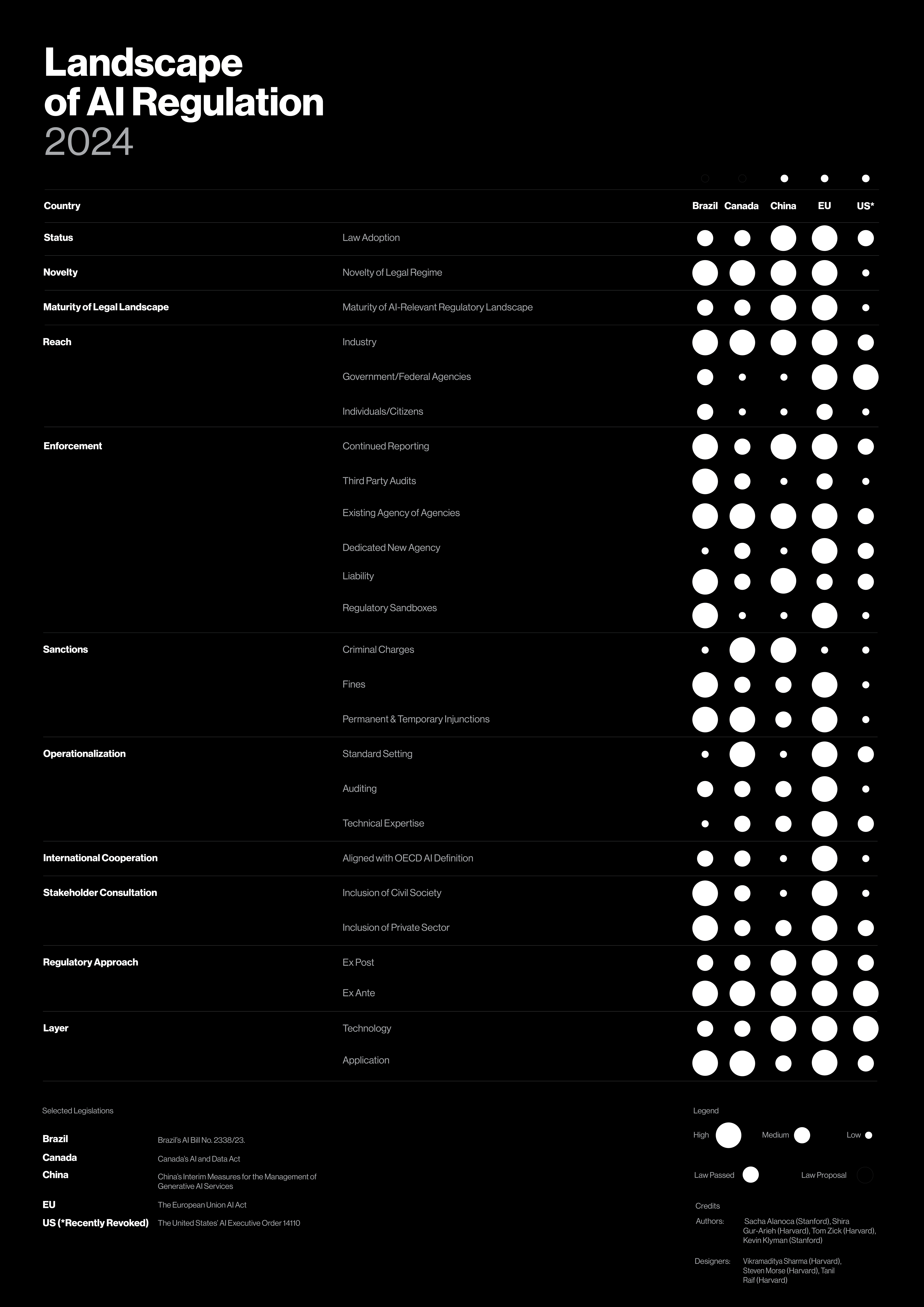}
  \caption{Taxonomy of the AI Regulation Landscape.}
  \label{fig:figure1}
\end{figure*}

\begin{figure*}[htbp]
  \centering
  % Try a fraction of \textheight. 0.7\textheight or 0.75\textheight is often a good start.
  % Adjust this value until the figure fits well with its caption.
  \includegraphics[height=0.95\textheight, width=\textwidth, keepaspectratio]{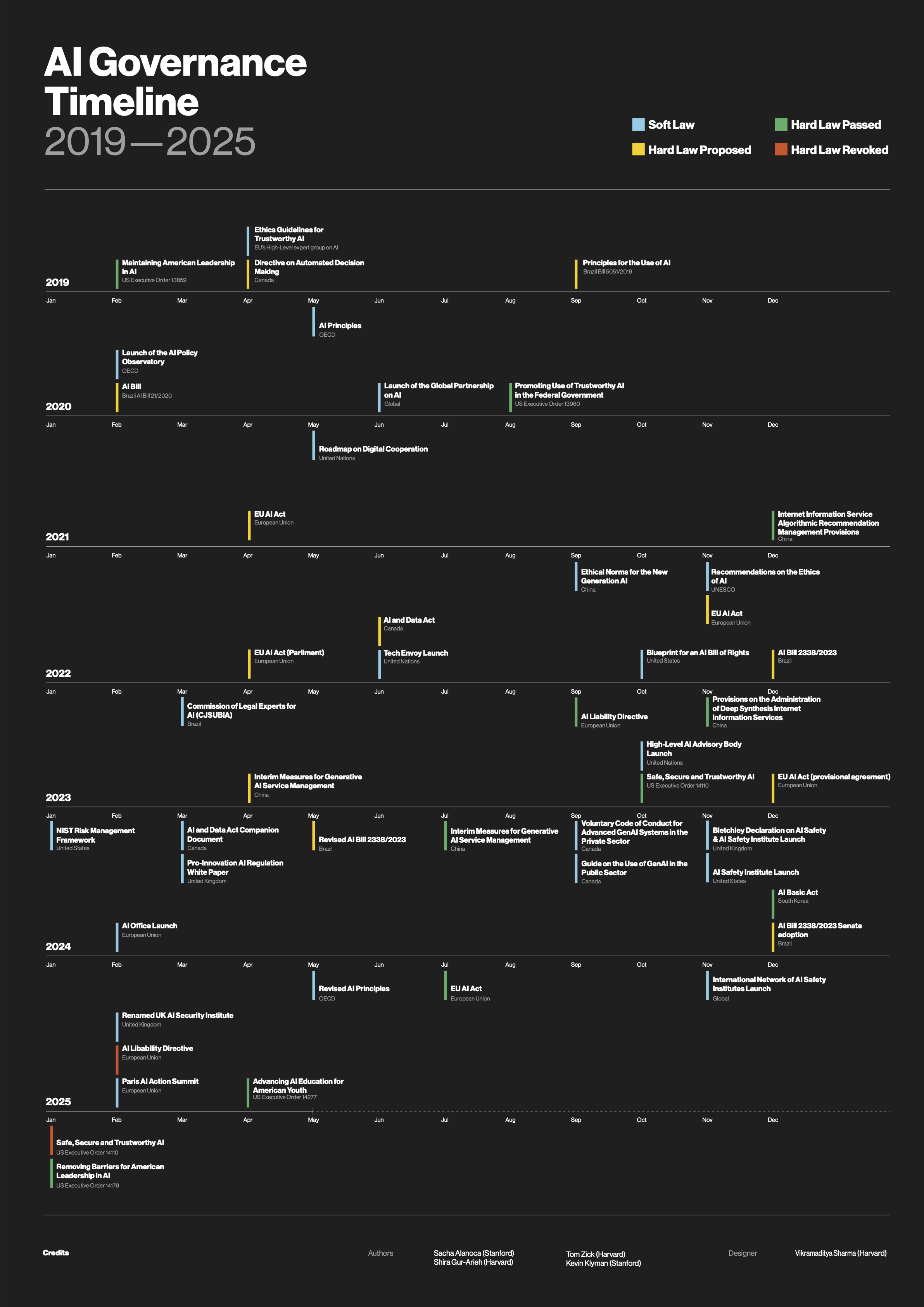}
  \caption{AI Governance Timeline.}
  \label{fig:figure2}
\end{figure*}

\section{Analysis and Results}
\label{sec:analysis_results}

\subsection{Different Stages of Development}
\label{subsec:stages_development}
A first step in navigating the landscape of AI regulation is distinguishing between adopted and pending legislation. The E.U., the U.S.\footnote{At the time of publishing this paper, the US EO 14110 has been revoked (\url{https://www.reuters.com/technology/artificial-intelligence/trump-revokes-biden-executive-order-addressing-ai-risks-2025-01-21/}) and Canada’s AIDA seems to have permanently been stalled (\url{https://montrealethics.ai/the-death-of-canadas-artificial-intelligence-and-data-act-what-happened-and-whats-next-for-ai-regulation-in-canada/}). While our current data visualization provides a snapshot of the current regulatory landscape, our taxonomy aims to be future proof to political and administration changes.}, and China have officially adopted legislative frameworks, while Brazil’s AI Bill 2338/2023 \cite{dataprivacybrazil2023legislation} and Canada’s AI and Data Act (AIDA) \cite{canadagovAIDACompanion} remain under consideration at the time this paper was drafted. Brazil’s bill has been approved by the Senate and is now being deliberated by the Chamber of Deputies. If adopted, it would mark a significant milestone as Latin America’s first comprehensive AI legislation. Canada’s AIDA, which was first presented in 2022, yields the risks of being permanently stalled if it is not being adopted ahead of the 2025 Federal elections. These prolonged timelines reflect the non-linear processes involved in regulating AI, often characterized by intense public deliberation and institutional negotiations, as seen in the E.U., Canada, and Brazil. A common pattern across these frameworks is the precedence of soft law initiatives over hard law efforts. For example, China’s Ethical Norms for New Generation AI (2021) and the U.S. Blueprint for an AI Bill of Rights (2022) laid the groundwork for subsequent legislation such as China’s Interim Measures for Generative AI Services (2023) \cite{pwc2023china} and the U.S. Executive Order 14110 (2023) \cite{usexecutive2023safe}. Similarly, the E.U.’s Ethics Guidelines for Trustworthy AI (2019) introduced foundational principles that informed the E.U. AI Act \cite{euparliament2024resolution}, thereafter, influencing regulatory approaches in other parts of the world. For example, while Brazil and Canada emphasize proportional obligations based on risk or impact levels across applications, similarly to the E.U., the U.S. and China focus on technology-based regulation (e.g, targeting powerful frontier) \cite{halim2023vectors, gsell2024regulating}. 

To summarize, three primary factors explain AI legislation asymmetrical timelines. First, the legislative depth and coverage of frameworks like the E.U. AI Act necessitates longer deliberations. The Act addresses AI systems and GPAIs comprehensively, offering a detailed classification of use cases by risk levels, which requires extensive cycles of rule design and negotiation. Second, the nature of the framework plays a significant role. For instance, executive orders like the U.S. EO 14110 bypass legislative deliberations entirely, allowing them to be issued directly by the White House. This contrasts sharply with the formal legislative processes seen in the E.U., Canada, and Brazil, which involve extensive institutional review. Finally, the degree of centralization in governance influences the speed of enactment. In China, the Cyberspace Administration of China (CAC), often referred to as a “Cyber Super-Regulator” which wields close to absolute regulatory powers, enabling the rapid enactment of AI-related measures \cite{wu2023interpret, toner2023howwill}.

\subsection{Different Levels of Maturity of the Digital Legal Landscape}
\label{subsec:maturity_landscape}
Emerging AI regulation stems from countries and regions with varying levels of maturity in their digital legal landscapes. As outlined by our taxonomy, at one end of the spectrum are the E.U. and China, which have achieved holistic coverage of emerging technologies through early regulatory efforts, including personal data privacy protections, content moderation, and copyright laws \cite{gsell2024regulating, bradford2020brussels, horsley2021howwill}. In the middle of the spectrum are Brazil and Canada, which have advanced soft law initiatives, including AI national strategies and comprehensive data privacy bills (e.g., Canada’s Privacy Law PIPEDA and Brazil’s General Data Protection Law), but lag in other aspects of tech-related regulation. At the other end is the U.S., which, despite its early AI policy efforts, has yet to develop comprehensive federal regulatory policies for emerging technologies, relying instead on tech-agnostic sectoral regulations and liability regimes \cite{levesque2024smoke}.

The E.U. represents one end of the spectrum with its extensive regulatory coverage \cite{halim2023vectors, gsell2024regulating, bradford2020brussels}. The E.U. AI Act, officially adopted in 2024, is considered the most holistic horizontal regulation of AI systems and has influenced global AI legislation. Embedded within a mature legal landscape, the AI Act complements landmark regulations such as the General Data Protection Regulation (GDPR) (2016), the Digital Services Act (DSA) and Digital Markets Act (DMA) (2022), and the Copyright Directive (2019). Additional frameworks, like the Cyber Resilience Act and Product Liability Directive (2024), further illustrate the E.U.’s precautionary approach. Critics, however, point to challenges in balancing safeguards with innovation and fragmented implementation due to varying national interpretations and enforcement capacities \cite{levesque2024smoke, gsell2024regulating}. 

China, often juxtaposed with the U.S. in the AI arms race narrative, is similarly mature to the E.U. in its digital legal landscape. Contrary to perceptions of unregulated tech dominance, China has enacted robust laws, spearheaded by its national internet regulation agency CAC. Key legislations include the Administrative Provisions on Algorithm Recommendation (2021), Deep Synthesis Management Provisions (2022), and Interim Measures for Generative AI Services (2023) \cite{gsell2024regulating, luo2024china, toner2023howwill}. These laws focus on national security and online information control, while complementing data protection and copyright regulations \cite{toner2023howwill, horsley2021howwill}. Despite demanding requirements, China remains a global leader in AI development, as exemplified by the recent release of its DeepSeek model and high volume of AI patents \cite{stanfordhai2024global}. While the intentions behind Chinese AI regulation may be motivated by a different set of values than Western ones, such as political dissent censorship, the co-existence of stringent regulation with cutting-edge innovation exemplifies that these two dynamics are not mutually exclusive. 

Brazil and Canada occupy the middle of the spectrum. Brazil began its initiatives in 2019 with its national AI strategy and first AI-focused bill, Bill 5051/2019. Subsequent efforts, such as Bill 2338/2023, adopted a risk-based framework inspired by the E.U. AI Act, categorizing AI systems into excessive risk, high-risk, and non-high-risk. If enacted, it will be Latin America’s first comprehensive AI law, complementing frameworks like Brazil’s General Data Protection Law and Internet Bill of Rights \cite{rsfBrazilianAI, dataprivacybrazil2023legislation, uechi2023brazilspath}. Canada introduced the AI and Data Act (AIDA) in 2022, mandating compliance for “high impact” AI systems. AIDA is part of a broader update of Canada’s federal data privacy law PIPEDA, initially adopted in 2000, and further complements the Directive on Automated Decision-Making. However, while Canada’s digital legal framework is mature along certain axes, it remains a work in progress with regards to AI given uncertainties surrounding the adoption of AIDA ahead of the 2025 federal elections \cite{gsell2024regulating, canadagovAIDACompanion}. 

At the other end of the spectrum is the U.S., which, despite its early leadership in AI policy \cite{stanfordhai2024global}, lacks federal tech-specific regulation. Preliminary efforts during the first Trump presidency, such as EO 13859 (2019), emphasized AI research and adoption by federal agencies. Initiatives under the Biden administration, like the Blueprint for an AI Bill of Rights (2022) and NIST AI Risk Management Framework (2023), continue this trajectory with a focus on equity and societal risks but remain soft law initiatives, reflecting reliance on sectoral regulation and liability frameworks \cite{levesque2024smoke, kaminski2023regulating}. With regards to data protection frameworks, the U.S. federal landscape features patchwork obligations, with the Children and Teens’ Online Privacy Protection Act (2025) regulating children’s data, the Gramm-Leach-Bliley Act (1999) and the Health Insurance Portability and Accountability Act (1996), regulating financial and health data respectively.

\subsection{Different Regulatory Approaches: Vertical and Horizontal Regulation}
\label{subsec:vertical_horizontal}
Countries adopt varying approaches to AI regulation. Some frameworks seek to regulate AI horizontally across a broad array of use cases, while others adopt a vertical focus by re-purposing sector-specific regulations, such as those targeting employment, housing, or transportation \cite{halim2023vectors}. While no framework is purely horizontal or vertical, our taxonomy indicates that each AI regulation tends to lean predominantly toward one approach. The E.U., Canada, and Brazil embrace predominantly a horizontal strategy, enacting requirements based on the risk-(E.U., Brazil) or impact-(Canada) classification of AI systems. Despite the semantic differences between risk thresholds (E.U., Brazil) and impact levels (Canada), these frameworks share a comprehensive scope. They filter AI applications into different tiers with proportional obligations. The E.U. AI Act exemplifies this horizontal approach, which is supported by complementary sectoral law such as the E.U. Medical Device Regulation (MDR). 

In contrast, the U.S. has historically adopted a vertical approach, relying on sector-specific law rather than developing novel, AI-focused regulation. This strategy emphasizes lighter-touch regulatory measures, voluntary guidance, and industry standards, as seen with the NIST AI Risk Management Framework (RMF) released in 2023. Sectoral laws have been extended to cover AI by various federal agencies. For instance, the Department of Justice and the Equal Employment Opportunity Commission (EEOC) have issued guidance on how AI-powered software used by employers could violate the Americans with Disabilities Act (ADA) \cite{ceimia2023framework}. Similarly, the Federal Trade Commission (FTC) has clarified how existing laws, such as the Fair Credit Reporting Act, apply to AI systems when they are used to deny individuals employment, housing, credit, insurance, or other benefits \cite{ftc2020using}. 

China represents a middle ground between horizontal and vertical regulation. It has implemented numerous hard law requirements targeting specific types of AI, such as recommender systems (2021) and generative algorithms (2022). However, these regulations are issued by the internet regulation agency CAC under the authority of broader primary statutes \cite{wu2023interpret, toner2023howwill}. This combination of sector-specific rules and overarching frameworks reflects China’s strategy between horizontal and vertical regulatory approaches. It is important to note that while frameworks may lean toward horizontal or vertical regulation, all incorporate a mixed approach to some degree. For example, as mentioned, the E.U. AI Act demonstrates the interplay between a horizontal framework and existing sectoral regulations, such as the MDR and the GDPR. In practice, national sectoral laws remain applicable alongside the AI Act, resulting in a multi-layered regulatory system. To prevent fragmented enforcement, European regulators must clarify the relationships among various tech-applicable laws, ensuring their interoperability \cite{halim2023vectors}. 

Each approach—horizontal and vertical—presents distinct advantages and challenges, which will be tested as enforcement begins and AI technologies evolve. Horizontal regulation offers a unified framework and broad protections across diverse applications but may lack flexibility for emerging innovations. Conversely, sectoral legislation allows for greater adaptability and interpretation in response to technological developments but may result in inconsistent enforcement across sectors and weak protections. Ultimately, a combination of these models will enable countries to align strategic priorities with local values.

\subsection{Different Regulatory Approaches: Ex Ante Versus Ex Post}
\label{subsec:exante_expost}
Ex ante versus ex post governance captures a central divide in how jurisdictions approach AI regulation. Building on Margot Kaminski’s discussion in \textit{Regulating the Risks of AI} \cite{kaminski2023regulating}, ex ante strategies prioritize risk regulation before AI technologies are widely deployed. They incorporate instruments such as mandatory impact assessments, licensing, or pre-market checks, mirroring how hazardous products—like chemicals or pharmaceuticals—undergo extensive review prior to approval. These measures strive to embed safety, privacy, and nondiscrimination requirements into AI’s design and deployment stages. By contrast, ex post regimes center on legal accountability after harm has occurred, relying on tort law, product liability, and consumer protection avenues that require injured parties to show causation and fault. Given AI’s opacity and complexity, post hoc redress is often challenging to achieve, making ex ante safeguards increasingly attractive. Still, no governance framework is purely one or the other; most jurisdictions blend both approaches. 

The E.U. AI Act epitomizes an ex ante emphasis, which enforces detailed, risk-based obligations—like pre-market conformity assessments—for high-risk applications. These obligations run in parallel with sectoral rules (e.g., health, finance) and strong ex post remedies, but the E.U.'s hallmark is an upfront preventive strategy that aims to “build in” consumer protections and rights compliance. Brazil has taken inspiration from the E.U.’s risk-based model: its proposed AI legislation mandates classification of AI systems, risk assessments, transparency requirements, and includes regulatory sandboxes. While it remains unclear whether an enforcement authority will have the power to penalize noncompliance, Brazil’s framework skews toward ex ante oversight, though it retains ex post elements as a backstop. Canada, similarly, couples ex ante obligations—such as impact assessments and an AI and Data Commissioner tasked with monitoring high-impact AI systems—with reliance on established privacy and consumer laws for ex post enforcement. 

Although the U.S. traditionally leans strongly toward ex post avenues, Biden’s Executive Order 14110 introduces significant ex ante features, particularly for high-compute AI models (e.g., dual-use foundation models operating above $10^{26}$ FLOPS). It mandates safety tests, risk assessments, and reporting requirements before deployment, aligning with preventive oversight strategies. The order also directs federal agencies to establish guidelines for AI safety and risk mitigation, signifying a notable shift toward proactive governance. Despite these ex ante measures, the U.S. continues to rely heavily on after-the-fact mechanisms: the Federal Trade Commission’s power to police unfair trade practices, consumer protection lawsuits, and state-level rulemaking. Meanwhile, China manifests a more centralized ex ante philosophy: its Interim Measures for Generative AI require developers to register models, conduct security assessments, and abide by content and national security guidelines before public release. The result is a stringent, top-down system that controls risk through broad government oversight, although China likewise preserves ex post penalties to ensure compliance.

\subsection{Different Regulatory Focus on the Technology or Application Level}
\label{subsec:tech_app_focus}
Emerging AI legislations vary in their regulatory focus. They can target AI systems either at the application level, depending on the context or use case, or at the technology level, based on the premise that certain technologies pose inherent risks. For instance, the U.S. Executive Order 14110 \cite{usexecutive2023safe} outlines that dual-use foundation models with training compute exceeding $10^{26}$ floating point operations per second (FLOPs) present de facto security risks, leading to reporting requirements \cite{usexecutive2023safe}. By contrast, frameworks such as the E.U. AI Act \cite{euparliament2024resolution} initially chose to regulate AI systems at the application level, segmenting use cases into different risk categories, each tied to proportional obligations \cite{euparliament2024resolution}. 

As highlighted in our taxonomy, the U.S. and China prioritize regulating technology types (e.g., Generative AI, dual-use foundation models) agnostic of its use case. On the other hand, the E.U., Canada, and Brazil initially regulated AI systems at the application-level (e.g., real-time biometric identification, deepfakes) before adding technology coverage to address general-purpose AI (GPAI) systems. Therefore, the E.U. AI Act exemplifies a hybrid approach by regulating AI systems at the application level through a four-risk-category framework while adding supplementary rules for GPAI technologies. In its initial first draft, released in April 2021, the E.U. AI Act focused on applications, banning use cases like social scoring or biometric identification in public spaces while imposing stringent requirements on high-risk systems \cite{euparliament2024resolution}. Minimal-risk cases were largely unregulated, recognizing that specific risks may arise independently of sectoral deployment. The rapid adoption of Generative AI during trilogue negotiations in late 2022, however, shifted the regulatory focus \cite{gsell2024regulating}. New provisions addressed GPAI models, requiring developers to document energy consumption, training datasets, and copyright policies. Similarly, Canada and Brazil updated their regulatory proposals to include provisions for GPAIs, aligning with the E.U.’s hybrid approach \cite{canadagovAIDACompanion, agenciacamara2021chamber, rsfBrazilianAI, uechi2023brazilspath}. 

At the technology level, the U.S. EO 14110 introduces binding obligations for private-sector entities either developing dual-use foundation models or controlling critical compute infrastructure. These include reporting requirements for systems exceeding $10^{26}$ FLOPS—slightly above current frontier models like GPT-4.1—and for computing clusters used in AI training \cite{bommasani2023decoding}. China’s regulations, such as Provisions on Deep Synthesis Internet Information Services (2022) and the Interim Measures for Generative AI Services (2023) target technologies, such as Generative AI, and particularly deepfakes. However, these laws also account for risks pertaining to misinformation and pornography \cite{wu2023interpret, toner2023howwill}. Such a wider angle illustrates an intent to primarily focus on technology-based regulation while addressing selected societal risks. Overall, hybrid AI regulatory approaches tend to be more flexible and comprehensive, considering both national security concerns (e.g., by targeting powerful frontier AI models) and societal ones (e.g., by tackling sensitive use cases in hiring, health care, or the public space). Rather than considering technology- and application-focused AI regulation as mutually exclusive, governments should see them as complementary and decide how to balance them depending on local priorities and capacities.

\subsection{Different Levels of Enforcement: Centralized or Decentralized Models}
\label{subsec:enforcement_models}
The enforcement of AI regulation spans a spectrum, from centralized to decentralized models, each offering distinct advantages and challenges. Largely centralized enforcement models, such as those employed in the E.U. and Canada, aim to ensure coherence and uniform oversight. In the E.U., the AI Office supports the European Commission and Member States in implementing the AI Act, playing an advisory role, assisting in market surveillance, overseeing conformity assessments for high-risk AI systems, and leading enforcement for GPAI ones \cite{euparliament2024resolution}. Similarly, Canada’s proposed AIDA introduces an AI and Data Commissioner with investigatory powers to request records, mandate audits, and suspend high-impact systems posing imminent risks \cite{levesque2024smoke, scassa2024oversight}. These centralized mechanisms facilitate consistent regulatory application and proactive risk management, particularly in addressing systemic issues like high-risk or GPAI systems. However, such frameworks face challenges, including resource limitations and uneven national enforcement capacities. At the extreme end of centralized enforcement lies China, whose regulatory framework is tightly coordinated by the CAC. The CAC ensures compliance through mandatory algorithmic labeling and strict monitoring of generative AI applications. While this top-down approach enables the rapid implementation of regulatory priorities, it raises concerns about accountable governance due to a lack of transparent checks and balances, which may in turn restrict civic freedoms. 

In contrast, decentralized enforcement models, as seen in the U.S. and Brazil, prioritize flexibility and leverage existing regulatory agencies. In the U.S., the AI Safety Institute (AISI), hosted by the National Institute of Standards and Technology (NIST), adopts a lighter-touch approach by operationalizing the NIST AI Risk Management Framework (RMF). The AISI also plays a role in global AI cooperation efforts, convening the first meeting of the International Network of AI Safety Institutes in November 2024 \cite{allen2024aisafety}. However, the AISI lacks regulatory authority to impose legally binding safety provisions on AI companies, relying instead on guidance regarding the misuse of dual-use foundation models and voluntary compliance to pre-deployment evaluation, for example. Similarly, Brazil employs a decentralized enforcement model through its National System for Artificial Intelligence Regulation and Governance (SIA). This network of sectoral authorities, including telecommunications and healthcare regulators, is coordinated by the National Data Protection Authority (ANPD). The SIA also integrates diverse stakeholders, such as the Permanent Regulatory Cooperation Council (CRIA) and the Committee of AI Experts and Scientists (CECIA), creating a multi-layered regulatory structure. While these decentralized models benefit from sector-specific expertise and adaptability, they risk fragmentation and reduced oversight in areas requiring cross-sectoral collaboration \cite{gsell2024regulating, uechi2023brazilspath}. For instance, Brazil’s SIA faces potential dilution of its authority, while the U.S. enforcement model’s reliance on self-monitoring raises accountability and safety concerns over the limits of self-regulation. 

Ultimately, the diversity of enforcement models demonstrates that effective AI regulation need not rely solely on centralized or decentralized structures. A hybrid approach, combining centralized coordination with decentralized execution, offers a promising middle ground. Such a model can harness the strengths of centralized oversight to address systemic risks while employing decentralized mechanisms for tailored enforcement. By harmonizing these approaches, jurisdictions can establish robust and adaptable AI regulations that foster both innovation and accountability in a rapidly evolving landscape.

\subsection{Different Degrees of Stakeholder Participation}
\label{subsec:stakeholder_participation}
The meaningful engagement of diverse stakeholders in the design, development, and enforcement of AI regulation is a critical aspect of participatory AI governance. However, it remains one of the most opaque and challenging dimensions to assess. It remains hard to evaluate the concrete inclusion of civic stakeholders’ input on legislative outcomes, beyond performative aspects of civic deliberation. In theory, participatory governance involves inclusive processes that incorporate perspectives from civil society, industry, academia, and trade unions, among others. In practice, the complexity and opacity of the legislative process, limited participation of external stakeholders, and self-funded involvement to join consultations, heighten the risks of regulatory capture. This translates in asymmetrical representation between public and private stakeholders across most AI regulations surveyed. Concretely, civil society input tends to be limited both during the consultation phase for drafting the law and in its implementation—such as when legal principles are translated into specific requirements by standardization bodies (e.g., CEN, CENELEC, and ETSI in the E.U.). 

The E.U. AI Act, Brazil's AI Bill, and Canada’s AIDA demonstrate efforts to enhance participatory governance through extended consultation periods involving external experts and stakeholders via presential workshops and online feedback \cite{levesque2024smoke, gsell2024regulating}. However, civil society organizations have criticized these processes for lacking transparency and inclusivity \cite{levesque2024smoke}. For example, Canada’s strong foundation in AI civic engagement, reflected by the 2017 Montreal Declaration on Responsible AI \cite{montrealdeclarationRESPONSIBLEAI}, has not consistently translated into legislative inclusion. AIDA’s introduction notably lacked civil society consultation, and of 253 stakeholders consulted, 216 were from the business sector, with minimal representation from marginalized groups \cite{clement2023preliminary}. In Brazil, significant asymmetry in lobbying power and financial means were showcased during the passing of mature digital legislations. For example, Google was accused of manipulating public opinions by promoting on its search engine entries criticizing the adoption of Bill 2630 to address online disinformation \cite{bridi2023brazil}. Such strong imbalances highlight the risk of regulatory capture, where industry lobbying dominates the legislative process. This can lead to the exclusion of certain AI systems from the 'high-risk' or 'high impact' categories, the strategic setting of legal requirements for Generative AI systems with compute thresholds just above those of currently released models, and the promotion of convenient interpretations of legal obligations during the standardization process. To mitigate risks of regulatory capture—and as emphasized in our methodology—there is an urgent need for greater transparency regarding the composition of external stakeholder groups, including the ratio of public to private inputs, and how these inputs are translated into legal outcomes.

Addressing these challenges requires strategic interventions to reconcile industry expertise with public interest, and the meaningful inclusion of civil society groups beyond cosmetic intervention. Strategies include funding civil society organizations to upskill their understanding of technical AI governance and its legal implications, while financially supporting the time spent sharing inputs, for example within standardization processes.

\section{Conclusion}
\label{sec:conclusion}
This study provides a comprehensive snapshot of the emerging AI regulation landscape, highlighting the diversity of regulatory approaches, enforcement models, and stakeholder participation across key jurisdictions. By developing a taxonomy that systematically evaluates these dimensions, we aim to clarify the scope and implications of AI regulation, mitigate risks of fragmentation, and facilitate international cooperation. The visualizations and analyses presented here are designed to democratize access to complex legislative processes, equipping a wide range of stakeholders with actionable insights. One of the chief goals of this paper is to delineate the boundaries between soft law and hard law efforts, distinguishing AI policy grounded in voluntary guidelines from AI regulation anchored in mandatory requirements. The loose semantic use of “AI regulation” often creates significant confusion, allowing some countries or regions to co-opt the terminology as a smokescreen for providing insufficient or superficial protections. This risks misleading the public about the true scope and strength of regulatory measures, thereby undermining trust and accountability in AI systems. 

Our taxonomy demonstrates that not all legislative frameworks sharing the term “AI regulation” are equivalent. For instance, the E.U. AI Act \cite{euparliament2024resolution} and Brazil’s AI Bill \cite{dataprivacybrazil2023legislation} stand out as comprehensive models that include more robust civil society participation and a comprehensive coverage of both national security and societal issues through hybrid regulatory approaches at the application and technology layer. Moreover, these legislations impose ex ante and ex post requirements, providing protections both before and after the deployment of AI systems. Their effectiveness remains to be tested during the implementation phase as well as with continuing technological developments. Moreover, this paper intends to demystify the common narrative that AI innovation and regulation are mutually exclusive. While the intentions behind Chinese AI regulation may be motivated by a different set of values—such as censorship of political dissent—the co-existence of stringent regulation with cutting-edge innovation exemplifies that these two dynamics are not inherently opposite. 

The relevance of this taxonomy is particularly critical in navigating an uncertain political landscape. The revocation of the EO 14110 in January 2025 further underscores the precarious nature of regulatory continuity in the absence of robust federal legislative frameworks. Despite such shifts, this taxonomy serves as a tool which can resist time and political change, to offer a common ground of understanding and inform future governance initiatives. As we live through a period of significant AI regulatory transitions—with new frameworks such as South Korea’s AI Basic Act being introduced and setbacks such as the withdrawal of Biden’s AI EO and the E.U. AI Liability Directive—a long-lasting taxonomy is needed to take stock of current changes. Future research should not only expand this taxonomy to include newly adopted legislations but also adapt it to examine novel regulatory developments such as state laws and closer analysis of meaningful external stakeholder participation.

\begin{acks}
We are thankful to the Harvard Berkman Klein Center for Internet and Society for their scholarly and financial support, and for recognizing the value of this project from its early stages. We are also grateful to our design team Vikramaditya Sharma, Steven Morse, and Tanil Raif for their patience and creativity in translating our research findings into accessible data visualizations. We are deeply grateful to the following experts for graciously sharing their time and expertise: Gabriele Mazzini, Filipe Medon, Giovana Carneiro, Isabella Ferrari, Graham Webster, Florence G’Sell, Maroussia Lévesque, Martha Minow, Mason Kortz, Manon Revel, Greg Leppert, Jessica Fjeld, Risto Uuk, Nicolas Moës, Caroline Meinhardt, Florian Martin-Bariteau, and Alvaro Morales. We also thank the members of the CWIP research group at Stanford University for their precious feedback: Angèle Christin, Tomás Guarna, and Rachel Bergmann.
\end{acks}

\bibliographystyle{ACM-Reference-Format}
\bibliography{references}

% --- START APPENDIX ---
\clearpage 

\appendix 

\addcontentsline{toc}{section}{Appendix} 

% Figure 1 of Appendix - WITH CUSTOM APPENDIX TITLE EMBEDDED
\begin{figure*}[t!] 
  \centering
  % --- Custom Appendix Title ---
  % This text will appear on the same page as the image, right before it.
  % We use a font style that might approximate a heading.
  % TAPS will handle final styling, so precise font matching isn't critical here.
  {\noindent\Large\sffamily\bfseries APPENDIX} % Using sans-serif bold, large. \noindent prevents indentation.
  \par\vspace{4ex} % Adds a bit of vertical space after this custom title.
  % --- End Custom Appendix Title ---
  
  \includegraphics[width=\textwidth, height=0.9\textheight, keepaspectratio]{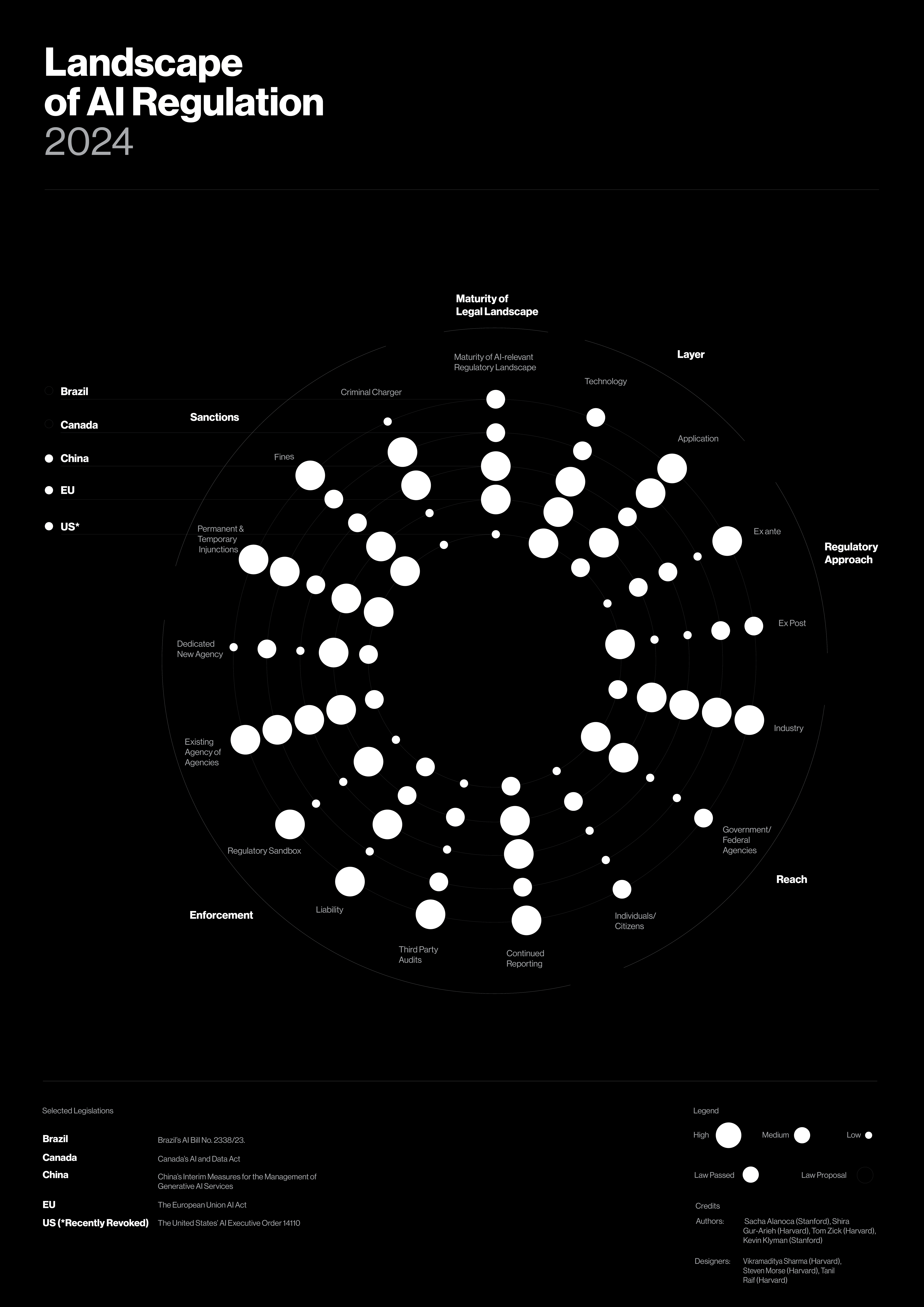}
  \caption{The Global Landscape of AI Regulation.}
  \label{fig:appendix_fig1} 
\end{figure*}

% --- TABLE PART 1 ---
\begin{table*}[htbp] % Use table* for full page width. 'htbp' for float placement.
    \centering
    \caption{Canada AI \& Data Act (AIDA) (Part 1 of 2)}
    \label{tab:aida-canada-p1}
    % \small % You might consider \small or \footnotesize if the table is too wide/long
    \begin{tabular}{p{0.18\linewidth} p{0.25\linewidth} p{0.50\linewidth}}
    \toprule
    \textbf{Category} & \textbf{Criteria} & \textbf{Metric} \\
    \midrule
    Status & Law adoption & Introduced in June 2022, the law remained permanently stalled after failing to advance ahead of the 2025 federal elections. \\
    \midrule
    Novelty & Novelty of legal regime & New law, even if it falls under Bill C-27 which proposes significant amendments to Canada's Personal Information Protection and Electronic Documents Act (PIPEDA). Bill C-27 would enact the Consumer Privacy Protection Act, the Personal Information and Data Protection Tribuna Act, and the AI \& Data Act (AIDA). \\
    \midrule
    Maturity of digital legal landscape & Maturity of AI-relevant regulatory landscape & AIDA is part of Bill C-27, the Digital Charter Implementation Act, which has not yet been passed. The regulatory landscape remains relatively premature, with current legislation leaving many essential components to be defined in future regulations. However, Canada has well-developed soft law initiatives: it was one of the first countries to release a national AI strategy in 2017 and is a founding member of the Global Partnership on AI (GPAI). \\
    \midrule
    Scope & Applies to Generative AI & Defines many related terms but not Generative AI, General-Purpose AI, or dual-use foundation models. \\
    \midrule
    \multirow{3}{*}[-1.5ex]{Reach} % The `*` width uses natural width of "Reach". Adjust [-1.5ex] for vertical alignment if needed.
    & Industry & AIDA is primarily targeted to industry stakeholders. In parallel, a Voluntary Code of Conduct on the Responsible Development and Management of Advanced Generative AI Systems in the Private Sector was released in September 2023. \\
    \cmidrule{2-3} % Rule only for columns 2-3
    & Government/ federal agencies & AIDA is not targeting government or federal agencies. However, in September 2023, a complimentary soft law was released: Canada's Guide on the Use of Generative AI in the Public Sector. \\
    \cmidrule{2-3}
    & Individuals/citizens & AIDA is not targeting directly individuals. \\
    \midrule
    \multirow{6}{*}[-7.5ex]{Enforcement} % Adjust vertical alignment offset as needed.
    & Continued reporting & An entity ``who is responsible for a high-impact system must, in accordance with the regulations and as soon as feasible, notify the Minister if the use of the system results or is likely to result in material harm.'' \\
    \cmidrule{2-3}
    & Third party audits & ``If the Minister has reasonable grounds to believe that a person has contravened any of sections 6 to 12 or an order made under section 13 or 14, the Minister may, by order, require that the person:(a) conduct an audit with respect to the possible contravention; or (b) engage the services of an independent auditor to conduct the audit.'' \\
    \cmidrule{2-3}
    & Existing agency or agencies & ``The Minister of Innovation, Science, and Industry would be responsible for administration and enforcement of all parts of the Act that do not involve prosecutable offences.'' \\
    \cmidrule{2-3}
    & Dedicated new agency & Not a new agency, but a new role: an AI and Data Commissioner, who would support the Minister in carrying out responsibilities. \\
    \cmidrule{2-3}
    & Regulatory Sandboxes & No regulatory sandboxes proposed. \\
    \cmidrule{2-3}
    & Liability & Not directly addressed. \\
    \bottomrule
    \end{tabular}
\end{table*}

% --- TABLE PART 2 ---
\begin{table*}[htbp] % Use table* for full page width. 'htbp' for float placement.
    \centering
    \caption{Canada AI \& Data Act (AIDA) (Part 2 of 2)} % Or use \caption*{} if you don't want it numbered/listed
    \label{tab:aida-canada-p2}
    % \small % You might consider \small or \footnotesize if the table is too wide/long
    \begin{tabular}{p{0.18\linewidth} p{0.25\linewidth} p{0.50\linewidth}}
    \toprule
    \textbf{Category} & \textbf{Criteria} & \textbf{Metric} \\
    \midrule
    \multirow{3}{*}[-4.5ex]{Sanctions}
    & Criminal Charges & AIDA creates three new criminal offences to directly prohibit and address specific behaviours of concern. The three offences are the following: (1) Knowingly possessing or using unlawfully obtained personal information to design, develop, use or make available for use an AI system. This could include knowingly using personal information obtained from a data breach to train an AI system; (2) Making an AI system available for use, knowing, or being reckless as to whether, it is likely to cause serious harm or substantial damage to property, where its use actually causes such harm or damage; (3) Making an AI system available for use with intent to defraud the public and to cause substantial economic loss to an individual, where its use actually causes that loss. \\
    \cmidrule{2-3}
    & Fines & Introduces Administrative Monetary Penalties (AMPs). AMPs are a flexible compliance tool that could be used directly by the regulator in response to any violation in order to encourage compliance with the obligations of the Act. While the Act allows for the creation of an AMPs regime, it would require regulations, and consultations, to come into force. \\
    \cmidrule{2-3}
    & Permanent \& Temporary Injunctions & ``The Minister may, by order, require that any person who is responsible for a high-impact system cease using it or making it available for use if the Minister has reasonable grounds to believe that the use of the system gives rise to a serious risk of imminent harm.'' \\
    \midrule
    \multirow{3}{*}[-3ex]{Operationalization}
    & Standard Setting & Intentions to set up a broad consultation which would lead to define ``the types of standards and certifications that should be considered in ensuring that AI systems meet the expectations of Canadians.'' \\
    \cmidrule{2-3}
    & Auditing & Not directly addressed. \\
    \cmidrule{2-3}
    & Technical Expertise & ``Commissioner could build a center of expertise in AI regulation'' and ``AIDA would mobilize external expertise in private sector, academia and civil society through: designation of external experts, use of AI audits, appointment of an advisory committee.'' \\
    \midrule
    International cooperation & International Cooperation & AIDA's companion document states that ``Canada has drawn from and will work together with international partners – such as the European Union (EU), the United Kingdom, and the United States (US) – to align approaches, in order to ensure that Canadians are protected globally and that Canadian firms can be recognized internationally as meeting robust standards.'' \\
    \midrule
    \multirow{2}{*}[-0.5ex]{Stakeholder consultation}
    & Inclusion of Civil Society & Intentions to include civil society: ``The AIDA will mobilize external expertise in the private sector, academia, and civil society to ensure that enforcement activities are conducted in the context of a rapidly developing environment.'' However, one preliminary analysis of Bill C-27’s stakeholder engagement found that, of the 253 stakeholders engaged, 216 belonged to the business sector and only a few were representatives of those might be affected and those at risk (Clement, 2023). \\
    \cmidrule{2-3}
    & Inclusion of Private Sector & One preliminary analysis of Bill C-27’s stakeholder engagement found that, of the 253 stakeholders engaged, 216 belonged to the business sector and only a few were representatives of those might be affected and those at risk (Clement, 2023). \\
    \midrule
    \multirow{2}{*}{Regulatory approach}
    & Ex Ante & High level of ex ante requirements. \\
    \cmidrule{2-3}
    & Ex Post & Medium level of ex post remedies. \\
    \midrule
    \multirow{2}{*}[-0.5ex]{Layer}
    & Technnology & Not directly targeting technology types such as Generative AI, GPAIs, or dual-use foundation models. \\ % Note: "Technnology" is as per CSV
    \cmidrule{2-3}
    & Application & Focus on ``high-impact'' AI systems that significantly affect individuals' health, safety and rights. \\
    \bottomrule
    \end{tabular}
\end{table*}

% --- EU AI ACT TABLE PART 1 ---
\begin{table*}[htbp]
    \centering
    \caption{EU AI Act (Part 1 of 2)}
    \label{tab:eu-ai-act-p1}
    % \small % You might consider \small or \footnotesize if the table is too wide/long
    \begin{tabular}{p{0.18\linewidth} p{0.25\linewidth} p{0.50\linewidth}}
    \toprule
    \textbf{Category} & \textbf{Criteria} & \textbf{Metric} \\
    \midrule
    Status & Law adoption & Officially presented by the European Commission in April 2021 and adopted by the EU in July 2024. \\
    \midrule
    Novelty & Novelty of legal regime & New law. \\
    \midrule
    Maturity of digital legal landscape & Maturity of AI-relevant regulatory landscape & Very mature landscape with the Global Data Protection Regime (GDPR), Digital Markets Act (DMA), Digital Service Act (DSA), and Chips Act. Challenge regarding potential overlaps (e.g., data governance requirements in GDPR vs. AI Act; general-purpose AI systems requirements in DSA vs. AI Act). \\
    \midrule
    Scope & Applies to Generative AI & Applies to general-purpose AI (GPAI) systems, models, and models with systemic risk. \\
    \midrule
    \multirow{3}{*}[-1.5ex]{Reach}
    & Industry & Law targets industry. \\
    \cmidrule{2-3}
    & Government/ federal agencies & Includes government and federal agencies. \\
    \cmidrule{2-3}
    & Individuals/citizens & Does not directly target individuals. \\
    \midrule
    \multirow{6}{*}[-7.5ex]{Enforcement}
    & Continued reporting & Serious AI incidents need to be reported to the AI Office and national competent authorities. \\
    \cmidrule{2-3}
    & Third party audits & The AI Act does not explicitly mention third-party audits, but its enforcement framework implies the emergence of a third-party ecosystem to assist companies with compliance, particularly regarding technical documentation, data processing practices, and risk management protocols. \\
    \cmidrule{2-3}
    & Existing agency or agencies & Market Surveillance Regulation Authorities are leveraged to enforce the AI Act, as well as the European Commission's DG Connect (which hosts the newly launched EU AI Office). \\
    \cmidrule{2-3}
    & Dedicated new agency & The European Commission established a new EU level regulator, the EU AI Office, which sits within DG CNECT. Launched in February 2024, the role of the AI office is to monitor, supervise, and enforce the AI Act requirements on general purpose AI (GPAI) models and systems across the 27 EU Member States. \\
    \cmidrule{2-3}
    & Regulatory Sandboxes & Art. 57: ``Member States shall ensure that their competent authorities establish at least one AI regulatory sandbox at national level. (...) Additional AI regulatory sandboxes at regional or local level, or established jointly with the competent authorities of other Member States may also be established.'' \\
    \cmidrule{2-3}
    & Liability & Not directly covered in the EU AI Act but in related acts. \\
    \midrule
    \multirow{3}{*}[-4.5ex]{Sanctions}
    & Criminal Charges & No mention of charges but of offenses and penalties in Annex IIa: ``Criminal offences will be charged on the following domains: Terrorism,Trafficking in human beings,Sexual exploitation of children and child pornography (...)``. \\
    \cmidrule{2-3}
    & Fines & Listed under penalties in Art. 5, 71 and 72a. ``Non-compliance with the prohibition of the AI practices referred to in Article 5 shall be subject to administrative fines of up to 35 000 000 EUR or, if the offender is a company, up to 7~\% of its total worldwide annual turnover for the preceding financial year, whichever is higher. (...) The Commission may impose on providers of general purpose AI models fines not exceeding 3~\% of its total worldwide turnover in the preceding financial year or 15 million EUR whichever is higher.'' \\
    \cmidrule{2-3}
    & Permanent \& Temporary Injunctions & Included. \\
    \bottomrule
    \end{tabular}
\end{table*}

% --- EU AI ACT TABLE PART 2 ---
\begin{table*}[htbp]
    \centering
    \caption{EU AI Act (Part 2 of 2)}
    \label{tab:eu-ai-act-p2}
    % \small
    \begin{tabular}{p{0.18\linewidth} p{0.25\linewidth} p{0.50\linewidth}}
    \toprule
    \textbf{Category} & \textbf{Criteria} & \textbf{Metric} \\
    \midrule
    \multirow{3}{*}[-3ex]{Operationalization}
    & Standard Setting & Three organisations are responsible for all EU standard-setting: CEN, CENELEC, and ETSI. CEN and CENELEC are taking the lead in creating standards in support of the Act. Alignment with international standard organizations such as ISO/IEC JTC 1 and its subcommittees like SC 42 AI. \\
    \cmidrule{2-3}
    & Auditing & ``Market surveillance authorities shall be granted access to the source code of the high-risk AI system upon a reasoned request and only when the following cumulative conditions are fulfilled: testing/auditing procedures and verifications based on the data and documentation provided by the provider have been exhausted or proved insufficient. The notified body shall carry out periodic audits to make sure that the provider maintains and applies the quality management system and shall provide the provider with an audit report. In the context of those audits, the notified body may carry out additional tests of the AI systems for which an EU technical documentation assessment certificate was issued.'' \\
    \cmidrule{2-3}
    & Technical Expertise & Present through three main channels: 1) The Joint Research Centre (JRC-Seville) (technical center of expertise); 2) The AI Office with technical recruits and the Scientific panel of independent experts (The Commission will select experts that demonstrate expertise, independence from AI developers, and the ability to execute their duties diligently, accurately, and objectively); 3) The European Centre for Algorithmic Transparency, launched in Sevilla. \\
    \midrule
    International cooperation & International Cooperation & AI Act definition of AI systems aligned with that of the OECD, and notably of the revised OECD AI Principles update of May 2024. \\
    \midrule
    \multirow{2}{*}[-0.5ex]{Stakeholder consultation}
    & Inclusion of Civil Society & Inclusion of civil society through: 1) The High-Level Group of AI Experts (AI HLG); 2) Feedback on the White Paper for Trustworthy AI; 3) Consultation/meetings request with EU Institutions. Lack of public data over degree of involvement of civil society and how inputs were translated into tangible legal outcomes. \\
    \cmidrule{2-3}
    & Inclusion of Private Sector & Inclusion of private sector through: 1) The High-Level Group of AI Experts (AI HLG); 2) Feedback on the White Paper for Trustworthy AI; 3) Consultation/meetings request with EU Institutions. Lack of public data over degree of involvement of private stakeholders and how inputs were translated into tangible legal outcomes. \\
    \midrule
    \multirow{2}{*}{Regulatory approach} % May not need vertical adjustment for 2 rows
    & Ex Ante & High level of ex ante requirements. \\
    \cmidrule{2-3}
    & Ex Post & Medium level of ex post remedies. \\
    \midrule
    \multirow{2}{*}[-0.5ex]{Layer}
    & Technology & Targeting GPAI systems, models, and models with systemic risk. \\
    \cmidrule{2-3}
    & Application & Risk classification of AI systems takes into account most sensitive applications. \\
    \bottomrule
    \end{tabular}
\end{table*}

% --- China Interim GenAI Measures - TABLE PART 1 ---
\begin{table*}[htbp]
    \centering
    \caption{China Interim GenAI Measures (Part 1 of 2)}
    \label{tab:china-genai-p1}
    % \small % You might consider \small or \footnotesize if the table is too wide/long
    \begin{tabular}{p{0.18\linewidth} p{0.25\linewidth} p{0.50\linewidth}}
    \toprule
    \textbf{Category} & \textbf{Criteria} & \textbf{Metric} \\
    \midrule
    Status & Law adoption & Draft introduced in April 2023 and law adopted in July 2023. \\
    \midrule
    Novelty & Novelty of legal regime & New law. \\
    \midrule
    Maturity of digital legal landscape & Maturity of AI-relevant regulatory landscape & China has an advanced digital legal landscape which includes the Administrative Provisions on Recommendation Algorithms in Internet-based Information Services (2021), the Deep Synthesis Provisions (2022), and a comprehensive data privacy law which was implemented in late 2021, the Personal Information Protection Law (PIPL). \\
    \midrule
    Scope & Applies to Generative AI & ``Article 22: The meanings of the following terms used in these Measures are: (1) ``Generative AI technology'' refers to models and relevant technologies that have the ability to generate content such as texts, images, audio, or video. (2) ``Generative AI service providers'' refers to organizations and individuals that use generative AI technology to provide generative AI services (including providing generative AI services through programmable interfaces and other means). (3) ``Generative AI service users'' refers to organizations and individuals that use generative AI services to generate content.'' \\
    \midrule
    \multirow{3}{*}[-1.5ex]{Reach}
    & Industry & Law targets industry. \\
    \cmidrule{2-3}
    & Government/ federal agencies & Does not fully target the public sector: ``Article 16: Based on their respective duties, departments such as for internet information, reform and development, education, science and technology, industry and informatization, public security, radio and television, and press and publication are to strengthen the management of generative AI services in accordance with law.'' \\
    \cmidrule{2-3}
    & Individuals/citizens & Does not fully target individuals: ``Article 18: Where users discover that generative AI services do not comply with laws, administrative regulations, or these Measures, they have the right to make a complaint or report to the relevant departments in charge.'' \\
    \midrule
    \multirow{6}{*}[-7.5ex]{Enforcement}
    & Continued reporting & Must submit safety assessment before license is granted. \\
    \cmidrule{2-3}
    & Third party audits & Not included. \\
    \cmidrule{2-3}
    & Existing agency or agencies & Law enacted and implemented by the Cyberspace Administration of China (CAC). \\
    \cmidrule{2-3}
    & Dedicated new agency & No new dedicated agency. \\
    \cmidrule{2-3}
    & Regulatory Sandboxes & No regulatory sandboxes. \\
    \cmidrule{2-3}
    & Liability & Where providers violate these Measures, penalties are to be given by the relevant regulatory departments in accordance with the provisions of the PRC Cybersecurity Law, the PRC Data Security Law, the PRC Law on the Protection of Personal Information, the PRC Law on Scientific and Technological Progress, and other such laws and administrative regulations; and where laws and administrative regulations are silent, the relevant departments in charge are to give warnings, circulate criticism, or order corrections in a set period of time on the basis of their duties, and if corrections are refused or the circumstances are serious, an order is to be given to suspend the provision of the related services. \\
    \bottomrule
    \end{tabular}
\end{table*}

% --- China Interim GenAI Measures - TABLE PART 2 ---
\begin{table*}[htbp]
    \centering
    \caption{China Interim GenAI Measures (Part 2 of 2)}
    \label{tab:china-genai-p2}
    % \small
    \begin{tabular}{p{0.18\linewidth} p{0.25\linewidth} p{0.50\linewidth}}
    \toprule
    \textbf{Category} & \textbf{Criteria} & \textbf{Metric} \\
    \midrule
    \multirow{3}{*}[-4.5ex]{Sanctions}
    & Criminal Charges & ``Article 21: Where violations of public security are constituted, a public security administrative sanction is lawfully given; where a crime is constituted, criminal responsibility is to be pursued in accordance with law.'' \\
    \cmidrule{2-3}
    & Fines & ``Article 21: Where providers violate these Measures, penalties are to be given by the relevant regulatory departments in accordance with the provisions of the PRC Cybersecurity Law, the PRC Data Security Law, the PRC Law on the Protection of Personal Information, the PRC Law on Scientific and Technological Progress, and other such laws and administrative regulations; and where laws and administrative regulations are silent, the relevant departments in charge are to give warnings, circulate criticism, or order corrections in a set period of time on the basis of their duties, and if corrections are refused or the circumstances are serious, an order is to be given to suspend the provision of the related services.'' \\
    \cmidrule{2-3}
    & Permanent \& Temporary Injunctions & ``Article 21: If corrections are refused or the circumstances are serious, an order is to be given to suspend the provision of the related services.'' \\
    \midrule
    \multirow{3}{*}[-3ex]{Operationalization}
    & Standard Setting & Not directly included. \\
    \cmidrule{2-3}
    & Auditing & Different wording around auditing but includes a corpus provisions in Basic Safety Requirements. \\
    \cmidrule{2-3}
    & Technical Expertise & No newly dedicated agency or mention of recruitments of technical experts per se, but evaluation provisons listed in Basic Safety Requirements will require technical expertise. \\
    \midrule
    International cooperation & International Cooperation & Not aligned. \\
    \midrule
    \multirow{2}{*}[-0.5ex]{Stakeholder consultation}
    & Inclusion of Civil Society & No known inclusion of civil society within the legal drafting process. \\
    \cmidrule{2-3}
    & Inclusion of Private Sector & There is some level of inclusion of private stakeholders, but the extent is unclear due to low public disclosure. \\
    \midrule
    \multirow{2}{*}{Regulatory approach}
    & Ex Ante & High level of ex ante requirements such as pre-registration. \\
    \cmidrule{2-3}
    & Ex Post & High level of ex post requirements such as having to monitor for bad uses and prevent them. \\
    \midrule
    \multirow{2}{*}[-0.5ex]{Layer}
    & Technology & Targets Generative AI services. \\
    \cmidrule{2-3}
    & Application & Targets certain contexts but predominantly focused on technology types. \\
    \bottomrule
    \end{tabular}
\end{table*}

% --- US AI EO 14110 - TABLE PART 1 ---
\begin{table*}[htbp]
    \centering
    \caption{US AI EO 14110 (Part 1 of 2)}
    \label{tab:us-eo14110-p1}
    % \small % You might consider \small or \footnotesize if the table is too wide/long
    \begin{tabular}{p{0.18\linewidth} p{0.25\linewidth} p{0.50\linewidth}}
    \toprule
    \textbf{Category} & \textbf{Criteria} & \textbf{Metric} \\
    \midrule
    Status & Law adoption & Released in October 2023 and revoked in January 2025. \\
    \midrule
    Novelty & Novelty of legal regime & Not a new law but an executive order. Operationalizes existing law and agencies with minor budgeting and personel changes. Empowers agencies to propulgate rules pertaining to AI using their existing delegated powers. \\
    \midrule
    Maturity of digital legal landscape & Maturity of AI-relevant regulatory landscape & No federal laws directly targeting technology or AI-focused regulation. However, there are existing regulatory actions taken by the FTC, lititgation under existing causes of action, and state laws (e.g., California Consumer Privacy Act (CCPA)). \\ % lititgation as per CSV
    \midrule
    Scope & Applies to Generative AI & Applies to dual-use foundation models, especially above certain computing capacities evaluated through the threshold of Floating-Point Operations Per Second (FLOPS) (e.g., above $10^{26}$ FLOPS for dual-use foundation models). \\
    \midrule
    \multirow{3}{*}[-1.5ex]{Reach}
    & Industry & Mainly targeted at directing federal agencies. There is only one direct requirement to industry within the EO (e.g., reporting for dual-use foundation models with $10^{26}$ FLOPS). \\
    \cmidrule{2-3}
    & Government/ federal agencies & Restrictions, procedures and calls for issuing rules on government use of AI. \\
    \cmidrule{2-3}
    & Individuals/citizens & Not primary target. \\
    \midrule
    \multirow{6}{*}[-7.5ex]{Enforcement}
    & Continued reporting & Yes, for example regarding dual-use foundation models above $10^{26}$ FLOPS. \\
    \cmidrule{2-3}
    & Third party audits & Not included. \\
    \cmidrule{2-3}
    & Existing agency or agencies & The EO mostly deals with the mandates of existing agencies with regards to AI; the only augmentations is the appointment of Chief AI officers to existing agencies. \\
    \cmidrule{2-3}
    & Dedicated new agency & Not a complete new agency with regulatory power, but launch in November 2023 of the AI Safety Institute (AISI) hosted by the National Institute of Standards and Technology (NIST). \\
    \cmidrule{2-3}
    & Regulatory Sandboxes & No regulatory sandboxes. \\
    \cmidrule{2-3}
    & Liability & Not directly covered. \\
    \bottomrule
    \end{tabular}
\end{table*}

% --- US AI EO 14110 - TABLE PART 2 ---
\begin{table*}[htbp]
    \centering
    \caption{US AI EO 14110 (Part 2 of 2)}
    \label{tab:us-eo14110-p2}
    % \small
    \begin{tabular}{p{0.18\linewidth} p{0.25\linewidth} p{0.50\linewidth}}
    \toprule
    \textbf{Category} & \textbf{Criteria} & \textbf{Metric} \\
    \midrule
    \multirow{3}{*}[-1.5ex]{Sanctions} % Adjusted vertical offset for 3 items
    & Criminal Charges & Not included: ``This directive does not apply to agencies’ civil or criminal enforcement authorities.'' \\
    \cmidrule{2-3}
    & Fines & Not directly provided by the EO: ``only within existing agency authority/agencies are encouraged to use their existing exisitng authority and rule making powers.'' \\ % exisitng as per CSV
    \cmidrule{2-3}
    & Permanent \& Temporary Injunctions & Similarly, not directly provided by the EO: ``only within existing agency authority/agencies are encouraged to use their existing exisitng authority and rule making powers.'' \\ % exisitng as per CSV
    \midrule
    \multirow{3}{*}[-3ex]{Operationalization}
    & Standard Setting & While standards are contemplated and delegated to the standardization agency NIST, the EO does not call for those standards to be met by hard law compliance. \\
    \cmidrule{2-3}
    & Auditing & Not addressed. \\
    \cmidrule{2-3}
    & Technical Expertise & Launch of the AI Safety Institute (AISI) to recruit technical experts, and of Chief AI officers to increase each agency's technical capacity. \\
    \midrule
    International cooperation & International Cooperation & Not aligned. \\
    \midrule
    \multirow{2}{*}[-0.5ex]{Stakeholder consultation}
    & Inclusion of Civil Society & Not directly included in the EO draft, but indirectly included through the NIST AI Safety Institute Consortium. \\
    \cmidrule{2-3}
    & Inclusion of Private Sector & Not directly included in the EO draft, but indirectly included through the NIST AI Safety Institute Consortium. \\
    \midrule
    \multirow{2}{*}{Regulatory approach}
    & Ex Ante & Requirements for dual-use foundation models above $10^{26}$ FLOPS, among other preconditions. \\
    \cmidrule{2-3}
    & Ex Post & Indirect liability elements. \\
    \midrule
    \multirow{2}{*}[-0.5ex]{Layer}
    & Technology & Targets dual-use foundation models. \\
    \cmidrule{2-3}
    & Application & Not targeting specific applications. \\
    \bottomrule
    \end{tabular}
\end{table*}

% --- Brazil AI Bill No. 2338/2023 - TABLE PART 1 ---
\begin{table*}[htbp]
    \centering
    \caption{Brazil AI Bill No. 2338/2023 (Part 1 of 2)}
    \label{tab:brazil-bill2338-p1}
    % \small % You might consider \small or \footnotesize if the table is too wide/long
    \begin{tabular}{p{0.18\linewidth} p{0.25\linewidth} p{0.50\linewidth}}
    \toprule
    \textbf{Category} & \textbf{Criteria} & \textbf{Metric} \\
    \midrule
    Status & Law adoption & In December 2024, Bill No. 2338/2023 was approved by the Brazilian Senate and now needs to be adopted by the Chamber of Deputies (as per May 2025). \\
    \midrule
    Novelty & Novelty of legal regime & New law. \\
    \midrule
    Maturity of digital legal landscape & Maturity of AI-relevant regulatory landscape & Complementing an emerging digital regulatory environment, such as the Internet Bill of Rights (MCI), Access to Information Law (LAI) and the General Data Protection Law (LGPD). \\
    \midrule
    Scope & Applies to Generative AI & Applies to general purpose AI (GPAIs) systems and generative AI systems. \\
    \midrule
    \multirow{3}{*}[-1.5ex]{Reach}
    & Industry & Law targets industry. \\
    \cmidrule{2-3}
    & Government/ federal agencies & Not primarily targeting government or federal agencies. \\
    \cmidrule{2-3}
    & Individuals/citizens & Not primarily targeting individuals. \\
    \midrule
    \multirow{6}{*}[-7.5ex]{Enforcement}
    & Continued reporting & Law requires the registration and documentation of the preliminary assessment conducted by the provider for accountability and reporting, duty filed by the competent authority. \\
    \cmidrule{2-3}
    & Third party audits & Audit to be completed by a competent authority, which could be an existing agency or new one such as a National System of AI Regulation and Governance. \\
    \cmidrule{2-3}
    & Existing agency or agencies & Leveraging existing agencies and ministries. However, more recent developments to potentially create a National System of AI Regulation and Governance agency. \\
    \cmidrule{2-3}
    & Dedicated new agency & Tentative but unclear plans to create a new authority (the National System of AI Regulation and Governance agency) to oversee the AI Bill implementation. \\
    \cmidrule{2-3}
    & Regulatory Sandboxes & Includes regulatory sandboxes. \\
    \cmidrule{2-3}
    & Liability & Includes a liability regime. \\
    \bottomrule
    \end{tabular}
\end{table*}

% --- Brazil AI Bill No. 2338/2023 - TABLE PART 2 ---
\begin{table*}[htbp]
    \centering
    \caption{Brazil AI Bill No. 2338/2023 (Part 2 of 2)}
    \label{tab:brazil-bill2338-p2}
    % \small
    \begin{tabular}{p{0.18\linewidth} p{0.25\linewidth} p{0.50\linewidth}}
    \toprule
    \textbf{Category} & \textbf{Criteria} & \textbf{Metric} \\
    \midrule
    \multirow{3}{*}[-1.5ex]{Sanctions} % Adjusted for 3 items
    & Criminal Charges & Not included. \\
    \cmidrule{2-3}
    & Fines & The penalties for non-compliance with the Bill can go up to BRL 50 million (approx. \$8M USD) or 2\% of the total turnover of the company. \\
    \cmidrule{2-3}
    & Permanent \& Temporary Injunctions & Included. \\
    \midrule
    \multirow{3}{*}[-3ex]{Operationalization}
    & Standard Setting & Not directly included within the law. \\
    \cmidrule{2-3}
    & Auditing & Audit to be completed by a competent authority, which could be an existing agency or new one such as a National System of AI Regulation and Governance. \\
    \cmidrule{2-3}
    & Technical Expertise & No new center of expertise or specific mention of technical talent recruitment mentioned in the latest bill proposal. \\
    \midrule
    International cooperation & International Cooperation & Aligned with some definitions set by international stakeholders: such as OECD's definition of AI and general-pupose AI sysems, and the EU AI Act's classification of AI systems. \\ % general-pupose sysems as per CSV
    \midrule
    \multirow{2}{*}[-0.5ex]{Stakeholder consultation}
    & Inclusion of Civil Society & Relatively to other AI regulations considered, higher level of inclusion of civil society in the consultation process. Unclear yet how civic inputs were translated into specific legal outcomes. \\
    \cmidrule{2-3}
    & Inclusion of Private Sector & Inclusion of private stakeholders during the consultation process. \\
    \midrule
    \multirow{2}{*}{Regulatory approach}
    & Ex Ante & High level of ex ante requirements. \\
    \cmidrule{2-3}
    & Ex Post & Medium level of ex post remedies. \\
    \midrule
    \multirow{2}{*}[-0.5ex]{Layer}
    & Technology & Targets, as a secondary regulatory layer, general-purpose AI systems. \\
    \cmidrule{2-3}
    & Application & Primarily targets context or application based AI and general-purpose AI systems. \\
    \bottomrule
    \end{tabular}
\end{table*}

%%END APPENDIX

\end{document}